# How essential are unstructured clinical narratives and information fusion to clinical trial recruitment?

Preethi Raghavan, MS, James L. Chen, MD, Eric Fosler-Lussier, PhD, Albert M. Lai, PhD
The Ohio State University, Columbus, OH

**Abstract**

Electronic health records capture patient information using structured controlled vocabularies and unstructured narrative text. While structured data typically encodes lab values, encounters and medication lists, unstructured data captures the physician's interpretation of the patient's condition, prognosis, and response to therapeutic intervention. In this paper, we demonstrate that information extraction from unstructured clinical narratives is essential to most clinical applications. We perform an empirical study to validate the argument and show that structured data alone is insufficient in resolving eligibility criteria for recruiting patients onto clinical trials for chronic lymphocytic leukemia (CLL) and prostate cancer. Unstructured data is essential to solving 59% of the CLL trial criteria and 77% of the prostate cancer trial criteria. More specifically, for resolving eligibility criteria with temporal constraints, we show the need for temporal reasoning and information integration with medical events within and across unstructured clinical narratives and structured data.

**Introduction**

The electronic health record (EHR) is a powerful repository of patient information that can be leveraged to build applications that benefit the clinical community such as clinical trial recruitment. Understanding and extracting information from EHRs enables reasoning with clinical variables and supports decision making.[1] EHRs record patient information both as data coded in structured format, as well as in the form of free text clinical narratives. Structured data typically contains demographics, patient birth and death information, lab values, encounters, and at times procedures and diagnosis lists. Unstructured data includes free text clinical narratives that correspond to different encounters generated at various points of time, including admission notes, history and physical reports, discharge summaries, radiology reports, and pathology reports.

Clinical trial recruitment may be semi-automated through information extraction from the EHR. Clinical trials have eligibility (inclusion and exclusion) criteria that describe characteristics and constraints that help determine if a patient qualifies for a trial. Typically, clinicians and trial recruitment coordinators identify potential clinical trial patients from characteristics described in their medical history and match them against the eligibility criteria for individual trials. This standard model of clinical trial enrollment is rife with errors. If the clinical staff is unfamiliar with a particular trial or if there are competing trials, an eligible patient may be overlooked. On the opposite extreme, the clinical trials staff may be asked to evaluate patients who are clearly not candidates.

This information mismatch has the potential to be streamlined. Generating automated queries corresponding to eligibility criteria and querying patient records from the EHR in order to identify qualifying patients provides an efficient and agnostic approach to clinical trials recruitment..The pertinent question then is whether structured data, being easier to automatically process and understand, has sufficient information to resolve these eligibility criteria, or if there is a need to extract and reason with medical concepts in unstructured clinical narratives. Researchers have often emphasized the importance of using clinical narratives for clinical decision support,[1] information retrieval,[2] question answering[1] and automated clinical trial recruitment.[4] Unstructured data in clinical narratives captures important decisions and relationships between medical concepts including causal (symptom caused disease), consequential (why a drug or treatment was administered) and temporal (symptom before disease/ treatment). Furthermore, Rosenbloom et al.[5] suggest that clinical notes containing naturalistic prose have been more accurate and reliable for identifying patients with given diseases, and more understandable to healthcare providers reviewing patient records. However, to the best of our knowledge, there are no prior empirical studies that evaluate the usefulness of structured vs. unstructured data considering their advantages and limitations for a clinical task.

In this paper, we study two datasets of structured and unstructured data with patients suffering from chronic lymphatic leukemia (CLL) and prostate cancer obtained from The Ohio State University Wexner Medical Center. Given a set of eligibility criteria from corresponding clinical trials, we evaluate the number of criteria that can be resolved using information from just the structured data and the number of criteria that require information extraction from and reasoning with unstructured clinical narratives and data. There are three main contributions of this work: 1) Empirical evaluation of the commonly assumed hypothesis that unstructured clinical text processing is required and that structured data alone is insufficient to accurately resolve eligibility criteria with the help of a clinical trial use case; 2) Demonstration of the need for cross-narrative temporal reasoning in solving certain temporal eligibility criteria; 3) Demonstration of the need for information fusion across structured and unstructured data in solving certain temporal eligibility criteria.

**Related Work**

The recent decade has seen considerable research in the natural language processing (NLP) of unstructured clinical text.[3,6-8] Fushman et al.[1] discuss how successful processing of clinical narratives is the key to overall success of automated clinical decision support systems. They stress the importance of medical concepts with the help of named entity recognition and learning relations between those named entities are important for better understanding clinical narrative text. Wang et al.[7] propose a framework for automated pharmacovigilance by applying NLP and association statistics on comprehensive unstructured clinical data from the EHR. They argue that previous algorithms have focused on coded and structured data, and therefore miss important clinical data relevant to this task. Medical NLP systems like Mayo's cTakes,[8] and MedLEE[7] have components specifically trained or designed for information extraction from clinical text.

There has been some work on modeling temporal knowledge in eligibility criteria to help effective clinical text processing.[9-10] Ross et al.[10] observe that temporal features were present in 40% of clinical trial criteria analyzed as part of their study, where the type of temporal expression in the criteria ranged from well-specified to loosely-specified. Similarly, there have been considerable efforts, including rule-based algorithms, temporal annotation of clinical corpora, and machine learning methods, towards learning temporal relations and generating timelines of medical events from unstructured clinical text.[11-13] Zhou et al.[11] extract temporal relations between medical events in discharge summaries. The CLEF project[12] uses a pairwise supervised classification approach to learn temporal relations between medical events within the same narrative. While temporal information has been studied in the intra-document context, there is not much prior work in cross-narrative temporal relation learning and information fusion. Carlo et al.[14] attempt to align medical problems in structured and unstructured EHR data using UMLS by studying the information overlap between structured ICD-9 diagnoses and unstructured discharge summaries. They conclude that this is a non-trivial task with the need for better methods to detect correlating structured and unstructured data before aligning them. Köpcke et al.[15] compare the eligibility criteria defined in trial protocols with patient data contained in the EHR in multi-site trials to determine the extent of available data compared with the eligibility criteria of randomly selected clinical trials. However, their study is restricted to structured data in the EHR.

In spite of the large body of recent work in processing structured and unstructured clinical narratives for temporal reasoning, and other NLP tasks, there are no prior studies that empirically evaluate the usefulness of structured vs. unstructured data for a clinical task. We perform an empirical analysis of CLL and prostate cancer patient records and evaluate the performance of structured and unstructured data in resolving clinical trial eligibility criteria. We specifically focus on criteria with temporal constraints and illustrate the need for unstructured clinical narrative analysis including cross-narrative temporal reasoning and information fusion.

**Patient Records and Clinical Trial Eligibility Criteria - Data Description**

The EHR data used in this study consists of medical records for 2060 CLL patients and 1808 prostate cancer patients. The CLL dataset contains 95 different types of unstructured reports including discharge summaries, history and physical reports, specialty reports such as wound care, operative notes, OB/GYN and psych evaluations, social work assessment, referral letters and progress notes. It also consists of radiology reports, pathology reports and cardiology reports. The total number of unstructured clinical narratives in the CLL dataset is 100704. The structured data consists of lab reports, procedures list, diagnoses list and encounters list.

The prostate cancer dataset consists of 2652 oncology reports, 1582 pathology reports, 6606 radiology reports as part of unstructured data. The structured data in this dataset includes a discharge medications list (30178 medications), laboratory values (939 values), and a medications list (141932 medications).

The clinical trials dataset consists of a set of top 100 clinical trials each, as defined by clinicaltrials.gov, for both CLL and prostate cancer.

**Methodology**

*Medical concept extraction* - We annotated the clinical trial criteria datasets with medical concepts, concept unique identifiers (CUIs) and semantic types using MetaMap.[14] We then extracted criteria containing the following semantic types: Disease or Syndrome, Laboratory or Test Result, Procedure, Sign or Symptom, and Pharmacological Substance. The criteria containing the Temporal Concept semantic type were labeled as temporal eligibility criteria. Similarly, we also annotated both patient datasets with medical concepts and the semantic types mentioned previously.

*Matching medical concepts across clinical trials and patient datasets* - In order to evaluate the degree of overlap between the clinical trials dataset and structured and unstructured data in the medical records dataset, we compute the *Match* between medical concepts across these datasets. The match functions are computed across the datasets as follows. 1) UMLS CUI Match where an exact CUI match is computed and 2) Phrase Match where we compute a match between medical concepts (textual fragment identified as the medical concept). Thus we have,

- *Match*(CUI in the trial dataset, CUI in structured data)
- *Match*(CUI in the trial dataset, CUI in unstructured data)
- *Match*(Phrase in the trial dataset, medical concept in the structured data)
- *Match*(Phrase in the trial dataset, medical concept in the unstructured data)

These match functions are computed for two levels of analysis - (1) medical concept-level, where we compare all the medical concepts in the trials dataset against the structured and unstructured data, and (2) eligibility criteria level, where we compare all the medical concepts in each criterion against the structured and unstructured data.

*The medical concept-level* match helps analyze the number and type of medical concepts typically found in the structured and unstructured datasets when solving clinical trial eligibility criteria. As shown in the algorithm below, we compute the *match* between all medical concepts in the clinical trials dataset and the structured data. If there are no matching concepts found in the structured data, we then compute a *match* with the unstructured data.

1. Calculate
   a. *Match*(CUI in the trial dataset, CUI in the structured data)
   b. *Match*(Phrase in the trial dataset, medical concept in the structured data)
2. If there are no match results from step 1, then calculate
   a. *Match*(CUI in the trial dataset, CUI in the unstructured data)
   b. *Match*(Phrase in the trial dataset, medical concept in the unstructured data)

*The eligibility criteria-level match* helps us analyze the number of criteria that can be solved by structured data, unstructured data or both. In order to evaluate the need for temporal reasoning and information fusion and constrain the number of eligibility criteria, we restricted the eligibility criteria-level analysis to criteria with temporal constraints. We compare each eligibility criterion against both structured data and unstructured data to determine if the concepts in the criterion require only structured data, only unstructured data or both datasets together for resolution, as shown in the algorithm below.

1. For all temporal eligibility criteria,
   a. For all medical concepts (from 1 to *n*) in the criterion
      i. $Match_1$(CUI in the criterion, CUI in the structured data) $\wedge$ ... $\wedge$ $(Match_n$(CUI in the criterion, CUI in the structured data)
      ii. $Match_1$(Phrase in the criterion, Phrase in the structured data) $\wedge$ ... $\wedge$ $(Match_n$(Phrase in the criterion, Phrase in the structured data)
2. If i OR ii returns *true,* then the criterion can be resolved by the structured data
3. Repeat step 1. by replacing "structured data" with "unstructured data"
   a. If step i OR ii returns *true,*
      i. the criterion can be resolved by the unstructured data
      ii. else the criterion can be cannot be resolved by a concept match across unstructured data
4. If in step 2, we get *true* for "structured" as well as "unstructured data",
   a. the criterion can be solved using *either* the structured or unstructured data.

The algorithm first compares all medical concepts in the eligibility criterion against all medical concepts in the structured data. If all the concepts in the criterion are found in the structured data, we conclude that the criterion may be resolved using the structured data. We then do a similar comparison for unstructured data and if all concepts in the criterion are found in the unstructured data, we conclude that the criterion may be resolved using the unstructured data.

*Information fusion* - In the case where all the concepts in the criterion are found in both the structured as well as the unstructured data, we conclude that the criterion can be solved using *either* the structured or the unstructured data. However, the criterion may also require both structured as well as unstructured data for resolution. Taking this into consideration, we define information fusion as follows.

Given medical concepts $\{m_1, ... ,m_n\}$ in a clinical trial criterion, if $S_k$ is a set of *k* concepts that match the structured data and $U_j$ is a set of *j* concepts that match the unstructured data, where $k, j>0$ and $k, j<n$. Now there are two possibilities.

1. $L = S_k \cap U_j$ is not empty. Here, *L* concepts match both structured and unstructured data.
2. $L = S_k \cap U_j$ is empty. Here, *L* concepts match the structured data and the remainder *j* concepts match the unstructured data. So $S_k$ and $U_j$ are disjoint.

*Temporal reasoning in unstructured data* - For subset of criteria that require unstructured data for resolution, we further analyze the temporal constraints in the criteria and attempt to answer the following questions. How many temporal constraints can be solved using coarse temporal reasoning within each clinical narrative? How many temporal constraints require more granular temporal ordering within each clinical narrative? How many temporal constraints require cross-narrative temporal reasoning?

In order to answer these questions, we run a CRF-based time-bin tagger[17] and learn to associate the medical events within each narrative with one of the coarse time-bins: "*way before admission, before admission, admission, after admission, discharge*". The time-bin tagger was trained on different patient records not part of this dataset. We also perform fine-grained temporally ordering by learning to rank medical concepts within a clinical narrative by their order of occurrence.[18] This gives us both a coarse ordering and a fine-grained ordering of medical concepts within each clinical narrative. These intra-narrative temporal orderings are then combined with the admission and discharge dates across narratives to generate a cross-document partially ordered timeline of medical concepts for each patient.

**Results**

The methodology is empirically evaluated by calculating the extent of *match* between the eligibility criteria dataset and the structured and unstructured datasets. The medical concept-level match results between the trials datasets, consisting of all eligibility criteria, and the structured and unstructured data are shown in Table 1. The CLL trials dataset has 2167 medical concepts and the prostate cancer dataset has 1019 medical concepts.

The CLL trials have a total of 1720 eligibility criteria, while the prostate cancer trials have 1325 eligibility criteria, containing diseases, procedures, tests, symptoms and medications. We observe that more than half of the medical concepts in the CLL and prostate patient data were only found in the unstructured data. The most frequent medical concept semantic types found in the unstructured datasets include Finding, Sign or Symptom, Disease or Syndrome, whereas the most frequent medical concept semantic type in the structured data includes Laboratory Test or Procedure, Pharmacological Substance and Disease or Syndrome. If the structured data has diagnoses and encounters lists, there tend to be overlapping Disease or Syndrome type concepts across the structured data and unstructured clinical narratives.

|  | CLL | | Prostate Cancer | |
| --- | --- | --- | --- | --- |
|  | CUI | Medical Concept | CUI | Medical Concept |
| Structured Data Match | 23% | 29% | 11% | 19% |
| Unstructured Data Match | 61% | 68% | 48% | 57% |

Table 1: Medical Concept-level Analysis on CLL and Prostate Cancer Trials and Patient Records

354 of the eligibility criteria in the CLL trials and 297 of the eligibility criteria in the prostate cancer trials have temporal constraints. Table 2 shows results from matching temporal clinical trial eligibility criteria against structured and unstructured data. In both patient datasets, matching the textual fragment identified as the medical concept gives us a higher *match* percentage than trying to match CUIs. Importantly, the dependence on unstructured data for resolution of temporal eligibility criteria is higher than structured data. There is especially a huge gap between the structured and unstructured data match in the case of prostate cancer, where structured data only contributes to the resolution of 9% of the criteria.

|  | CLL | | Prostate Cancer | |
| --- | --- | --- | --- | --- |
|  | CUI | Medical Concept | CUI | Medical Concept |
| Structured Data Match | 35% | 37% | 9% | 9% |
| Unstructured Data Match | 53% | 59% | 75% | 77% |

Table 2: Eligibility Criteria-level Analysis on CLL and Prostate Cancer Trials and Patient Records

|  | CLL | Prostate Cancer |
| --- | --- | --- |
| Cross-Narrative Temporal Reasoning | 33% | 35% |
| Information fusion $L = S_k \cap U_j$ is not empty | 24% | 3% |
| Information fusion $L = S_k \cap U_j$ is empty | 17% | 1% |

Table 3: Eligibility Criteria that require Cross-narrative Temporal Reasoning and Information Fusion for resolution

We observed that from the temporal criteria requiring unstructured data for resolution, frequently intra-narrative temporal reasoning was sufficient for resolving temporal constraints. The learned time-bins, along with the admission and discharge dates on each narrative, were useful in assigning medical concepts to coarse time-periods and in resolving 41% of the eligibility criteria that required an unstructured data match. For instance, the constraint, "patients with a *distant history* (*greater than 6 months before* study entry) of venous thromboembolic disease are eligible", requires mapping of venous thromboembolic disease to a time-bin *way before time*. Whereas "clinically significant bleeding event *within the last 3 months*, unrelated to trauma, or underlying condition that would be expected to result in a bleeding diathesis" required fine-grained temporal ordering of medical concepts.

Further, as shown in Table 3, from the criteria that required unstructured data for resolution, 33% and 35% required cross-narrative temporal reasoning in the CLL and prostate cancer dataset respectively. A criteria such as, "fever > 100.5°F for 2 weeks without evidence of infection", requires extracting the fact that fever lasted for 2 weeks by examining multiple mentions of fever across history and physical reports and discharge summaries to determine when fever started and stopped. This additionally requires the ability to perform coreference resolution across clinical narratives.[19] Criteria requiring information from both structured and unstructured data (information fusion) were determined based on the presence of the medical concepts in the criteria across these data sources. For instance, "if they have achieved stable blood pressure (bp) on a regimen of over 2 drugs after *6-8 weeks* of therapy." The value of bp can be obtained from the structured data, however the nuanced relationship information about the drug regimen that was prescribed to stabilize bp, along with its time duration, requires time-bin learning and cross-narrative temporal reasoning.

We observed that while a large percentage of CLL criteria required fusion, the lower number of prostate cancer criteria is mainly due to limited structured data available for prostate cancer.

**Discussion**

We studied two datasets of patients – CLL and prostate cancer – and evaluated the usefulness of structured vs. unstructured data in recruiting for corresponding clinical trials. We observed that the type of structured data, its granularity, and the information available vary across patient datasets. While the CLL patient dataset has detailed structured data in the form of diagnoses lists, encounters list, procedures and lab values, the prostate cancer dataset has limited structured data mostly consisting of medication lists and lab values. More fundamentally, the data heterogeneity reflects the underlying tumor heterogeneity at multiple levels. These levels include: (1) patient referral patterns (2) patterns of disease treatment (3) and differences in disease stages. At The OSU James Cancer Hospital, the majority of prostate cancer patients tend to be referrals from community oncologists or urologists after failure of first and second line therapies. In contrast, CLL patients are mostly evaluated from time of diagnosis and thus their entire case history is within the OSU system. Secondly, laboratory values for prostate cancer patients are often drawn at their local laboratory and subsequently faxed to their oncologist at OSU. These labs are not directly accessible and are found in the unstructured component of the medical record. In stark contrast, CLL labs are nearly universally drawn at OSU.

These tumor type differences would help explain our findings that prostate cancer requires the use of the unstructured data more frequently. The end result is that prior treatment history for prostate cancer patients who are seen at a later stage will have their disease course and treatment course summarized in the unstructured narrative. CLL patients are captured at an earlier stage and therefore their disease course and treatment history is more easily obtained from the structured text. This tumor type heterogeneity is reflected in the diagnosis codes that are available. In the case of CLL, these codes are useful in checking eligibility criteria that check for the presence or absence of a medical condition can be resolved easily from the structured data using these lists. In case of prostate cancer, this data is not as complete.

Tumor heterogeneity aside, structured data may also fail if the medical concept is at a finer level of granularity than what is required for an exact match. In such cases, examining the unstructured data for additional information, or additional processing to check for related higher level concepts for medical events in the structured data may help better resolve the eligibility criteria.

**Conclusion**

We performed an empirical evaluation of clinical trial eligibility criteria resolution using structured and unstructured patient datasets from CLL and prostate cancer. We observed that unstructured data is essential to resolving eligibility criteria in 59% of the CLL trial criteria and 77% of the prostate cancer trials. We also demonstrated the need for cross-document temporal relation learning and information fusion across structured and unstructured data sources. Although structured data is useful in resolving certain criteria, it is limited by information granularity and structured data type. Thus, structured data is best used for first pass filtering of EHR data in eliminating a criterion based on the presence or absence of a certain lab test or diagnoses, prior to a more nuanced second pass using unstructured data. Moreover, improving the coverage of the structured data in the EHR would improve its ability to be used as a clinical trial recruitment tool.

**Acknowledgements**

The project described was supported by Award Number Grant R01LM011116 from the National Library of Medicine. The content is solely the responsibility of the authors and does not necessarily represent the official views of the National Library of Medicine or the National Institutes of Health.